\documentclass[twocolumn,prl,showpacs,preprintnumbers,amsmath,amssymb,letter]{revtex4-1} 
\usepackage{graphicx}
\usepackage{dcolumn}
\usepackage{bm}
\begin{document}
\title{Single-Ion Atomic Clock with $3\times10^{-18}$ Systematic Uncertainty}
\author{N.~Huntemann}
\email{nils.huntemann@ptb.de}
\author{C.~Sanner}
\author{B.~Lipphardt}
\author{Chr.~Tamm}
\author{E.~Peik}
\affiliation{Physikalisch-Technische Bundesanstalt, Bundesallee 100, 38116 Braunschweig, Germany}

\date{\today}
\begin{abstract}
We experimentally investigate an optical frequency standard based on the $^2S_{1/2} (F=0)\to {}^2F_{7/2} (F=3)$ electric octupole (\textit{E}3) transition of a single trapped $^{171}$Yb$^+$ ion. For the spectroscopy of this strongly forbidden transition, we utilize a Ramsey-type excitation scheme that provides immunity to probe-induced frequency shifts. The cancellation of these shifts is controlled by interleaved single-pulse Rabi spectroscopy which reduces the related relative frequency uncertainty to $1.1\times 10^{-18}$. To determine the frequency shift due to thermal radiation emitted by the ion's environment, we measure the static scalar differential polarizability of the \textit{E}3 transition as $0.888(16)\times 10^{-40}$~J m$^2$/V$^2$ and a dynamic correction $\eta(300~\text{K})=-0.0015(7)$. This reduces the uncertainty due to thermal radiation to $1.8\times 10^{-18}$. The residual motion of the ion yields the largest contribution $(2.1\times 10^{-18})$ to the total systematic relative uncertainty of the clock of $3.2\times 10^{-18}$. 

\end{abstract}


\maketitle

Today's most advanced atomic clocks use optical reference transitions of single ions in radio-frequency traps or ensembles of neutral atoms confined in an optical lattice \cite{Ludlow2015}. For both types of optical clocks, relative systematic frequency uncertainties below $10^{-17}$ have been reported. These systems employ a $^1S_0 \to {}^3P_0$ transition in either $^{27}$Al$^+$ \cite{Chou2010b} or neutral $^{87}$Sr \cite{Nicholson2015,Ushijima2015}. The frequency uncertainty achieved for $^{27}$Al$^+$ is limited by the residual motion of the sympathetically cooled ion \cite{Chou2010b}. In the case of $^{87}$Sr, for which so far the smallest systematic uncertainty has been achieved \cite{Nicholson2015}, the relatively large Stark shift resulting from thermal radiation of the atoms' environment needs to be either suppressed by cryocooling \cite{Ushijima2015} or corrected with high accuracy \cite{Bloom2014,Beloy2014}. In contrast to these systems,  the  $^2S_{1/2} (F=0)\to {}^2F_{7/2} (F=3)$ electric octupole (\textit{E}3) transition of $^{171}$Yb$^+$ offers advantages due to its small sensitivity to electric and magnetic fields and the ion's large mass implying small residual motion. The technical simplicity of trapping and laser cooling of Yb$^+$ has stimulated its application in various experiments, see e.g.\ Refs.~\cite{Olmschenk2009a,Zipkes2010,Timoney2011}. Furthermore, $^{171}$Yb$^+$ has the important advantage of a second narrow linewidth transition, the $^2S_{1/2} (F=0)\to {}^2D_{3/2}(F=2)$ electric quadrupole (\textit{E}2) transition, which also can serve as the reference of an optical frequency standard \cite{Tamm2014,Godun2014}. The significantly higher sensitivity of the \textit{E}2 transition to electric and magnetic fields permits diagnosis of field-induced shifts of the \textit{E}3 transition frequency on a magnified scale. 

Exploiting these advantages of the Yb$^+$ system, we report in this Letter an analysis of systematic frequency shifts of the \textit{E}3 transition yielding a total uncertainty of $3.2\times 10^{-18}$, which is more than an order of magnitude smaller than previously published values \cite{Godun2014,Huntemann2014}. We determine the static scalar differential polarizability of the \textit{E}3 transition with high accuracy, strongly reducing the related uncertainty that dominated in previous work. Together with an evaluation of the thermal radiation in our ion trap \cite{Dolezal2015}, it enables a correction of the shift caused by thermal radiation at room temperature with an uncertainty of $1.8\times 10^{-18}$. Addressing the large shift of the \textit{E}3 transition frequency by the probe light, we introduce an interrogation scheme \cite{Yudin2010,Huntemann2012b} that cancels the shift with $1.1\times 10^{-18}$ uncertainty. The high accuracy of this frequency standard makes it now possible to exploit the high sensitivity of the $^{171}$Yb$^+$ $^2F_{7/2}$ state energy in searches for variations of fundamental constants \cite{Godun2014,Huntemann2014}, violations of Lorentz invariance \cite{Dzuba2015} and ultralight scalar dark matter \cite{Arvanitaki2015,Stadnik2015}.

In our experimental setup \cite{Tamm2009,Huntemann2012a} a single $^{171}$Yb$^+$ ion is confined in a radio-frequency Paul trap and laser cooled on the $^2S_{1/2}\to {}^2P_{1/2}$ electric-dipole transition at 370~nm, while repump lasers at 935~nm and 760~nm prevent population trapping in the metastable $^2D_{3/2}$ and $^2F_{7/2}$ states. 

\begin{figure}
\includegraphics[width=\columnwidth]{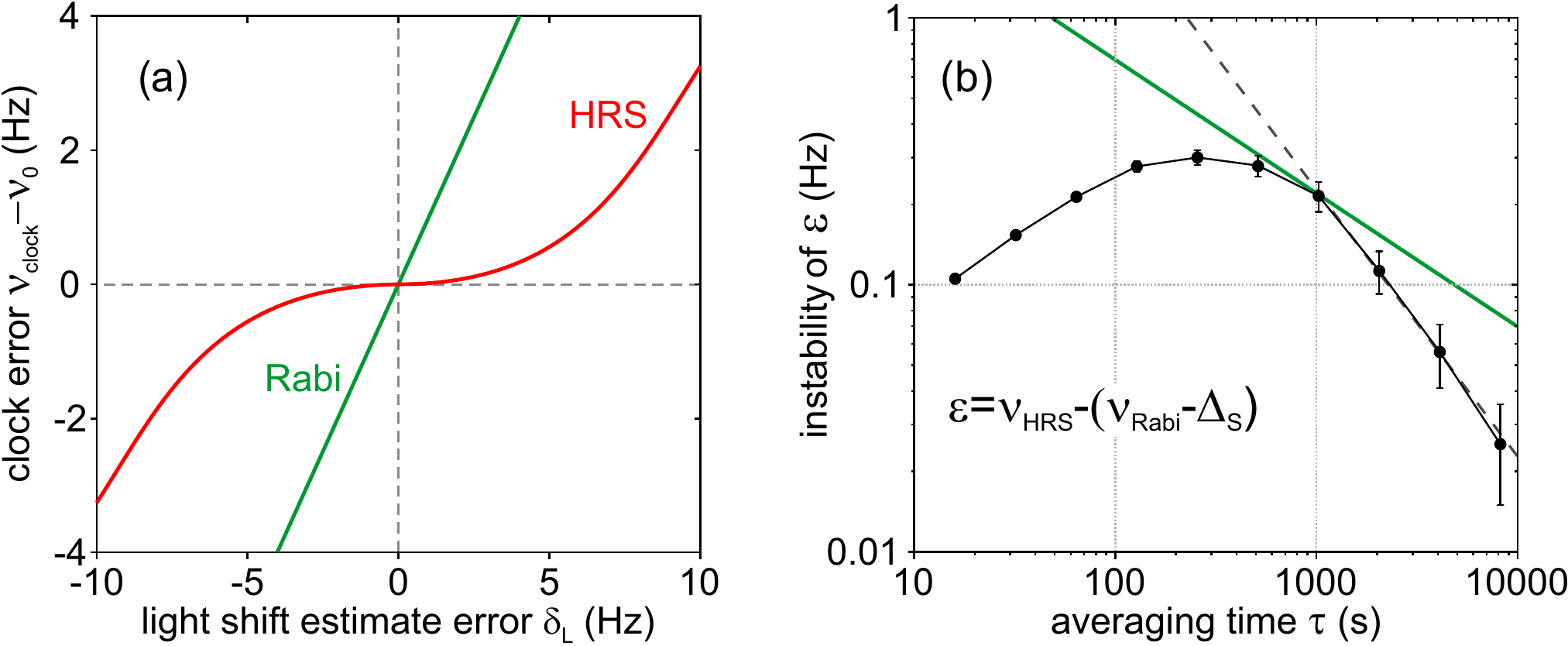}
\caption{\label{HRS} (a) Error of the Yb$^+$ clock frequency $\nu_\text{clock}$  in the realization of the unperturbed transition frequency $\nu_0$ as a function of an error $\delta_L$ in the estimate of the light shift for Rabi and for hyper-Ramsey spectroscopy (HRS). Here, Ramsey pulses of 30.5~ms and a free evolution period of 122~ms are assumed according to the experimental conditions. The very different sensitivities of $\nu_\text{clock}$ to $\delta_L$ allow one to engage a servo that uses the difference $\varepsilon$ between $\nu_\text{clock}$ obtained for Rabi spectroscopy and HRS as the discriminator signal. In (b) the instability (Allan deviation) of experimental $\varepsilon$ data is shown that follows $230$~Hz/$\tau(\text{s})$ (dashed line) for $\tau \geq 1000$~s. 
The green solid line indicates the expected quantum projection noise limited combined instability of the $\nu_\text{Rabi}$ and $\nu_\text{HRS}$ measurements of $7$~Hz/$\sqrt{\tau(\text{s})}$.
}
\end{figure}

While the small natural linewidth of the \textit{E}3 transition allows one to obtain very high resolution, the correspondingly small oscillator strength implies that a relatively high probe light intensity is required for excitation. Consequently, a significant light shift is induced via nonresonant coupling to higher-lying levels. For typical experimental parameters, the shift exceeds the observed Fourier-limited linewidth. Because of this light shift $\Delta_L$, an optical clock based on conventional Rabi spectroscopy of the \textit{E}3 transition cannot directly access the unperturbed transition frequency $\nu_0$ but will rather lock the probe laser to a frequency $\nu_\text{Rabi}=\nu_0+\Delta_L$. Additional measurements with altered light intensities can be used to provide an estimate of $\Delta_L$ \cite{Huntemann2012a} and a light shift correction $-\Delta_S$ can be applied to obtain $\nu_\text{clock}=\nu_\text{Rabi}-\Delta_S$. However, a light shift estimate error $\delta_L=(\Delta_L-\Delta_S)$ will map one to one to a clock error $\nu_\text{clock}-\nu_0$, as depicted in Fig.~\ref{HRS} (a).

The situation is different for Ramsey spectroscopy where a significant fraction of frequency information is accumulated during an interaction-free state evolution time. To maintain a resonant drive of the clock transition 
(i.e.\ to optimally initialize the atomic superposition state), one has to apply a frequency step $\Delta_S$ to the probe light during the interaction periods in order to reach the light-shifted resonance frequency $\nu_0+\Delta_L$ \cite{Taichenachev2009}. The step frequency $\Delta_S$ is readily obtainable from Rabi spectroscopy. If Ramsey spectroscopy is modified in this way, an error $\delta_L$ translates linearly to an error in $\nu_\text{clock}$ but via a reduced prefactor proportional to the interaction time fraction. Hyper-Ramsey spectroscopy (HRS) as introduced in Refs.~\cite{Yudin2010,Huntemann2012b} can further reduce the error by removing the linear sensitivity of $\nu_\text{clock}$ for small light shift estimate errors [see Fig.~\ref{HRS}~(a)]. Heating of the ion's motion during the probe period, however, reduces the effective pulse area of the second Ramsey pulse, degrading the cancellation of the linear dependence of $\nu_\text{HRS}$ on $\delta_L$ \cite{Tabatchikova2013,TaichenachevPC}. Under our experimental conditions, we estimate a residual linear slope $\partial\nu_\text{HRS}/\partial\delta_L=0.07$ at $\delta_L=0$. In order to confine $\delta_L$ to the region where the HRS clock error remains small, use can be made of the sensitivity of $\nu_\text{clock}$ to $\delta_L$ in Rabi spectroscopy. A feedback loop that employs the difference $\varepsilon=\nu_\text{HRS} -(\nu_\text{Rabi}-\Delta_S)$, as determined with HRS and Rabi excitations, as the discriminator signal can be used to steer $\Delta_S$ so that $|\delta_L|$ approaches zero. Figure~\ref{HRS}~(b) shows the instability (Allan deviation) of $\varepsilon$, i.e., the error signal of the operating $\Delta_S$ control loop. In this way, $\delta_L=0$ is realized for $\tau > 1000$~s with a statistical uncertainty given by the combined statistical uncertainties of the $\nu_\text{HRS}$ and $\nu_\text{Rabi}$ measurements, which are predominantly determined by quantum projection noise  \cite{Huntemann2012a}. For the measurement shown in Figure~\ref{HRS} (b), this combined instability is approximately given by 7~Hz/$\sqrt{\tau(\text{s})}$. Because of the small sensitivity of $\nu_\text{HRS}$ to $\delta_L$, the contribution to the $\nu_\text{HRS}$ instability arising from $\delta_L$ fluctuations is significantly smaller than the intrinsic quantum projection noise. The observed clock instability, i.e., that of $\nu_\text{HRS}$, amounts to 3~Hz/$\sqrt{\tau(\text{s})}$ corresponding to a fractional frequency instability of $5\times 10^{-15}/\sqrt{\tau(\mathrm{s})}$. 

Although the combination of interrogation schemes converts the uncertainty due to the light shift into a predominantly statistical contribution, 
systematic shifts can be caused by drifts of the light shift and by a difference of the shifts present during the Rabi and the HRS interrogations. Slow variations of the light shift corresponding to a drift of $\Delta_L$ in the range of $50~\mu$Hz/s are typically observed during our measurements. With a servo time constant of 200~s for $\Delta_S$, the resulting servo error of $\delta_L$ is $10$~mHz. This error could be avoided by using a drift-compensating second-order integrating servo algorithm \cite{Peik2006}. The main reason for differences in the frequency shifts present during the interrogation pulses are transient thermal effects of the crystal of the AOM that shapes the pulses. The resulting phase variations (AOM chirp) were investigated with a digital phase analyzer and found to lead to a frequency difference of less than 1~mHz \cite{Kazda2015}. Beam pointing and focusing variations induced by different crystal temperatures were found to lead to light shift differences between the pulses of less than 0.2~mHz. The combination of these systematic effects and the residual sensitivity $\partial\nu_\text{HRS}/\partial\delta_L=0.07$ yields a probe-light-related fractional uncertainty of $1.1\times 10^{-18}$. This is a reduction by more than an order of magnitude compared to previous \textit{E}3 clock realizations, 
where real-time extrapolation to zero probe laser intensity was used to cancel the light shift 
\cite{Huntemann2012a,Godun2014}.

\begin{figure}
\includegraphics[width=.85\columnwidth]{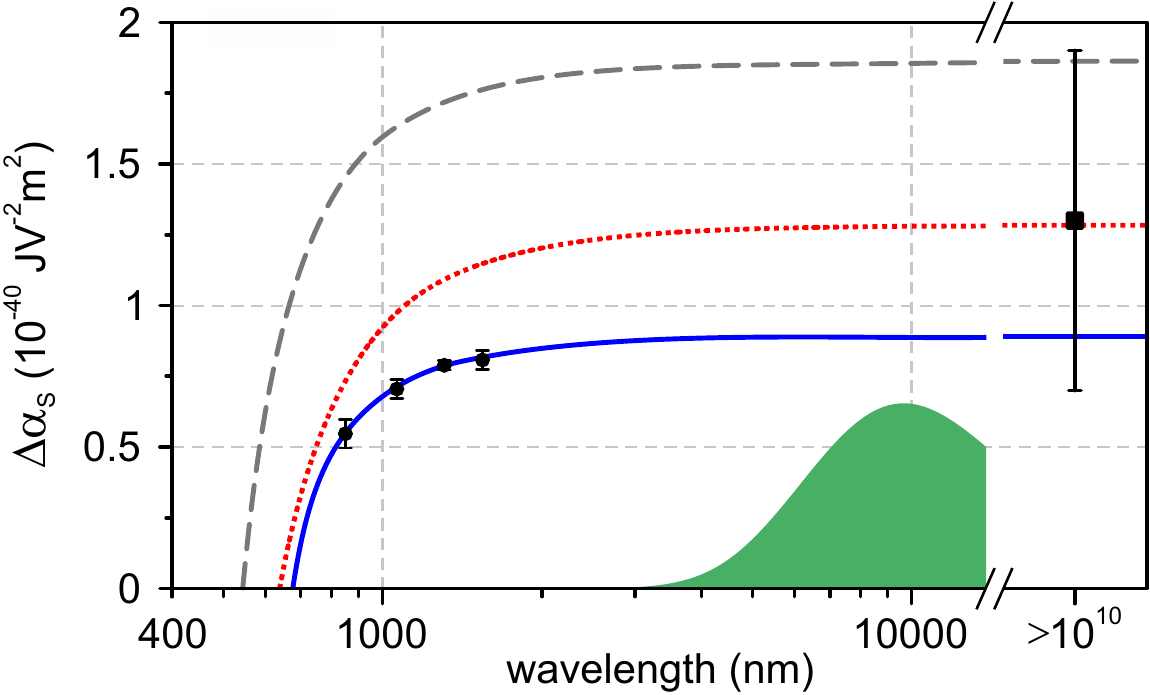}
\caption{\label{Pol} Scalar electric differential polarizability $\Delta\alpha_S$ of the $^2S_{1/2} (F=0)\to {}^2F_{7/2}(F=3)$ transition as a function of the wavelength of the perturbing radiation. The dashed and dotted lines are the results of calculations using theoretically predicted oscillator strengths, with the latter corrected by measured lifetimes. The square data point indicates the result of a previous measurement \cite{Huntemann2012a} and the filled circles indicate data obtained with near-infared laser radiation. The solid blue line is the result of a least-squares fit to the data (see text). The green shaded area shows the spectral distribution of room temperature blackbody radiation.}
\end{figure}

The largest shift of the transition frequency in our experiment is caused by the Stark shift induced by the thermal radiation emitted by the ion's environment. The temperature distribution of the various components of our ion trap is sufficiently homogeneous to approximate the electric field perturbing the transition frequency by a blackbody radiation (BBR) field at an effective temperature $T$ \cite{Dolezal2015}. Under this approximation, only the difference $\Delta \alpha_s=\alpha_s(e)-\alpha_s(g)$ of the scalar polarizabilities of the excited and the ground state is needed to evaluate the BBR shift. For the  $^{171}$Yb$^+$ \textit{E}3 transition, none of the transitions that contribute to $\Delta\alpha_S$ significantly overlap with the BBR spectrum at room temperature (see Fig.~\ref{Pol}), so that the BBR shift $\Delta \nu_\text{BBR}(T)$ can be expressed using the static scalar differential polarizability $\Delta \alpha_s^\text{dc}$ as 
\begin{equation}
\Delta \nu_\text{BBR}(T)=-\frac{1}{2h}\Delta \alpha_s^\text{dc} \langle E^2(T)\rangle (1+\eta(T)). 
\label{eq:TBBR}
\end{equation}
Here, $h$ is Planck's constant, $\langle E^2(T)\rangle$ is the mean-squared electric field inside the blackbody at temperature $T$, and $\eta(T)$ corrects for the variation of $\Delta\alpha_s$ in the range of the BBR spectrum and scales to first order quadratically with $T$ \cite{Porsev2006}. 

In a first experimental investigation $\Delta \alpha_s^\text{dc}$ had been determined to be $1.3(6)\times 10^{-40}$~J m$^2$/V$^2$ for the \textit{E}3 transition \cite{Huntemann2012a}. The large uncertainty dominated the uncertainty of optical clocks that use the \textit{E}3 transition as the reference \cite{Godun2014,Huntemann2014}. Several theoretical studies \cite{Biemont1998,Lea2006,Porsev2012} investigated complex electronic structure of Yb$^+$ and attempted to derive a value for $\Delta \alpha_s^\text{dc}$, but so far no approach has achieved sufficiently low uncertainties. The theoretical results can be corrected through measured state lifetimes \cite{Lea2006}, which changes $\Delta \alpha_s^\text{dc}$ by about 40\% (see Fig.~\ref{Pol}). Since the polarizabilities of the excited and the ground state are nearly equal, small corrections have a large effect on $\Delta \alpha_s^\text{dc}$.

\begin{figure}
\includegraphics[width=\columnwidth]{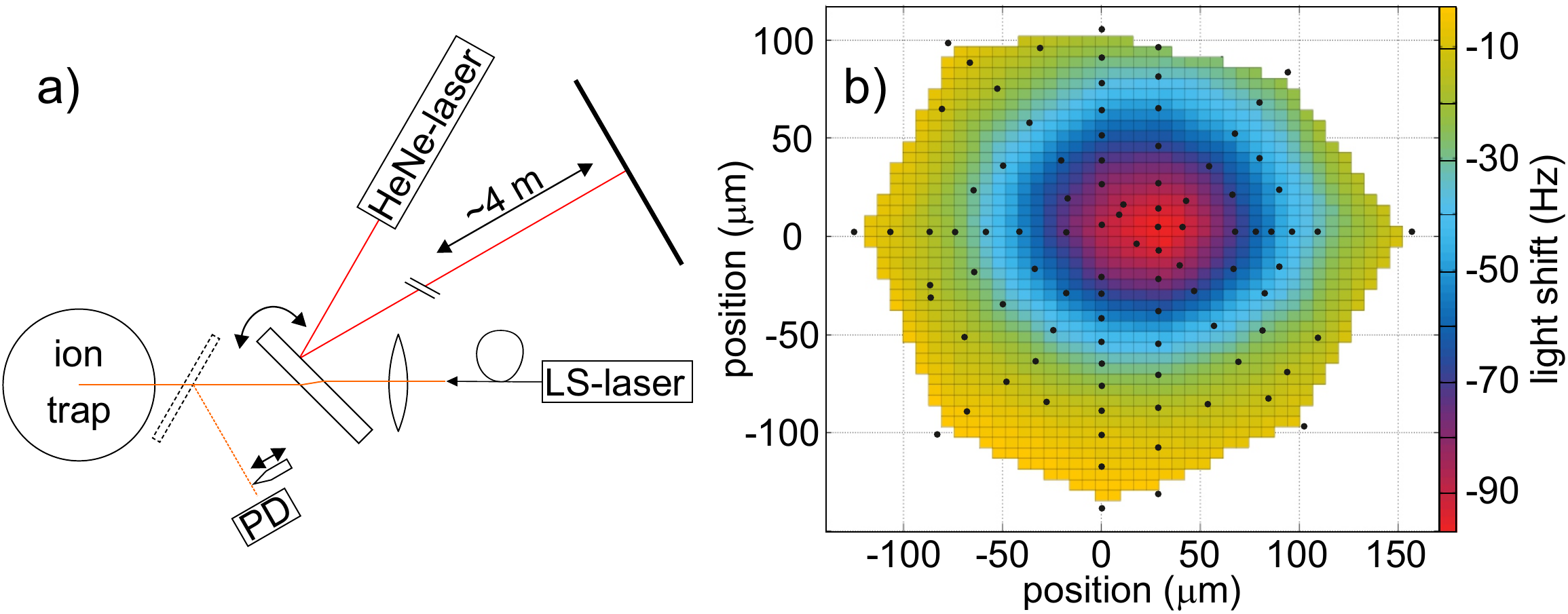}
\caption{\label{LSMeas} (a) Schematic of the setup used to determine the light shift profile. The mirror shown with dashed lines was installed after recording the profile to calibrate the position of the HeNe laser on the screen to a displacement of the beam waist position of the light-shifting laser (LS laser) at the position of the ion using a knife edge. In (b) a light shift profile induced by the LS laser at 1.5~$\mu$m is depicted. The black dots correspond to the measurement positions and the surface plot is the result of a linear interpolation between these points.}
\end{figure}

Since all transitions that contribute to the electric polarizability are at wavelengths below 380~nm, $\Delta\alpha_s^\text{dc}$ can be investigated using near-infrared (NIR) laser radiation \cite{Rosenband2006}. Note that the very small matrix element of the $^2F_{7/2}\to {}^2D_{5/2}$ transition at 3.43~$\mu$m changes $\Delta \nu_\text{BBR} (300~\text{K})$ negligibly by less than 0.1\% \cite{Taylor1999}. We account for the residual spectral dependence of $\Delta\alpha_s(\lambda)$ by performing light shift measurements at various wavelengths. The output beam profiles of the selected lasers at 852, 1064, 131, and 1554~nm were cleaned by single-mode fibers, yielding an output power of about 100~mW focused to a beam waist diameter of about 100~$\mu$m at the center of the trap. Figure~\ref{LSMeas}~(a) sketches the experimental setup. The linear polarization of the laser light was aligned to minimize losses at the windows of the vacuum enclosure that are mounted close to Brewster's angle. By averaging the induced light shift over three mutually orthogonal orientations of the magnetic field, its scalar part is isolated. The measurement was performed using the interleaved servo technique \cite{Tamm2009}, with the NIR laser light alternately applied and blocked during the interrogation periods. We measure the applied optical power and determine the relative intensity distribution at the position of the ion through light shift measurements for various displacements of the beam. The optical power is assumed to be the average of the power values measured in front of and behind the trap. The power meter was calibrated with an uncertainty of 0.5\%. The relative optical power at each beam displacement was monitored using a linear photodetector. This power monitoring was continuous during the recording at 1310~nm. For all other wavelengths, potential power fluctuations cause an increased uncertainty of the optical power of 3\%. Beam displacement was achieved by tilting a 3.1~mm thick glass plate around Brewster's angle in front of the ion trap. The reflection of a pointer laser on a screen at a distance of approximately 4~m monitored the tilt angle. The relation between tilt angle and beam displacement was established as shown in Fig.~\ref{LSMeas}(a). The relative uncertainty of this calibration is about 0.3\% for both coordinates. At each NIR wavelength at least two light shift profiles were recorded. A typical profile is shown in Fig.~\ref{LSMeas} (b). The measured light shift distribution is fitted by an intensity distribution composed of elliptical TEM$_{0,0}$, TEM$_{1,0}$, and TEM$_{0,1}$ Gauss-Hermite modes with relative residuals well below 1\%. Dividing the spatially integrated light shift profile by the measured optical power yields $\Delta \alpha_s(\lambda)$. 

In order to extrapolate $\Delta \alpha_s(\lambda)$ from our experimental data points to the spectral range of blackbody radiation at room temperature, we use a function composed of a static contribution $\Delta \alpha_s^\text{dc}$ and a wavelength-dependent term describing the dynamic character of $\Delta \alpha_s(\lambda)$ caused by a single resonance at $\lambda_0$:
\begin{equation}
\Delta \alpha_s(\lambda)=\Delta \alpha_s^\text{dc}-\frac{C}{\lambda^2 - \lambda_0^2 }.\label{Deltaalpha}
\end{equation}
For $\Delta \alpha_s^\text{dc}=-C/\lambda_0^2$, this expression resembles the response of a two-level system to a far-detuned polarizing field \cite{Grimm2000}. To better account for a manifold of contributing transitions, we keep $\Delta \alpha_s^\text{dc}$, $C$, and $\lambda_0$ as independent fit parameters. 
As a test of the model, we use it to fit the polarizability obtained from calculated oscillator strengths (see Fig.~\ref{Pol}) in the range of 850~nm to 1550~nm. The fit result reproduces the calculated $\Delta \alpha_s^\text{dc}$ value to better than 0.2\% and we assume this as the uncertainty of our model. A fit to our experimental data yields $\Delta \alpha_s^\text{dc}=0.888(16)\times 10^{-40}$~J m$^2$/V$^2$, where the largest contribution to the combined uncertainty results from the optical power measurement. The fit also gives the value of the dynamic correction as $\eta(300~\text{K})=-0.0015(7)$. 

To obtain the BBR shift with determined values of $\Delta \alpha_s^\text{dc}$ and $\eta$, one needs to know the effective temperature $T$ of the thermal radiation at the location of the ion. A combination of finite element modeling, infrared camera and temperature sensor measurements reveals a temperature rise of $2.1(1.1)$~K above room temperature, mostly caused by dielectric losses in the insulators of the trap assembly \cite{Dolezal2015}. From Eq.~(\ref{eq:TBBR}) the BBR shift can be calculated as $-45.3$~mHz, corresponding to a relative shift of $-70.5(1.8)\times 10^{-18}$. Here, the uncertainties of temperature and polarizability contribute approximately equally to the combined uncertainty. 

\renewcommand{\arraystretch}{1.2}
\begin{table}
\caption{\label{UncertBudget} Fractional frequency shifts $\delta\nu/\nu_0~(10^{-18})$ and related relative uncertainties $u/\nu_0~(10^{-18})$ in the realization of the unperturbed $^2S_{1/2} (F=0)\to {}^2F_{7/2} (F=3)$ transition frequency $\nu_0$ of a single trapped $^{171}$Yb$^+$ ion.}
\begin{tabular} { l c c}
 \hline\hline 
 Effect                           & $\delta\nu/\nu_0~(10^{-18})$                & $u/\nu_0~(10^{-18})$  \\ \hline 
 Second-order Doppler shift       & $-3.7$                                      & $2.1$   \\
 Blackbody radiation shift        & $-70.5$                                     & $1.8$   \\
 Probe light related shift        & $0$                                         & $1.1$   \\
 Second-order Zeeman shift        & $-40.4$                                     & $0.6$   \\ 
 Quadratic dc Stark shift         & $-1.2$                                      & $0.6$   \\  
 Background gas collisions 				& $0$                                         & $0.5$   \\
 Servo error                      & $0$                                         & $0.5$   \\  
 Quadrupole shift                 & $0$                                         & $0.3$   \\ \hline
 Total			                      &	$-115.8$	                                  & $3.2$ 	\\ 
\hline\hline
\end{tabular}
\end{table}

Table~\ref{UncertBudget} summarizes frequency shifts and the related uncertainty contributions of the Yb$^+$ single-ion \textit{E}3 clock. The magnetic field of $3.58(2)~\mu$T present during the interrogation is alternatingly applied at one of three orientations that are mutually orthogonal with an uncertainty of $1^\circ$ in order to suppress tensorial shifts \cite{Itano2000}. The listed second-order Zeeman shift and the uncertainty due to the quadrupole shift are calculated under these conditions. The uncertainty associated with collisions with the background gas is estimated using a model based on phase changing Langevin collisions \cite{Rosenband2008}. Another small uncertainty contribution results from the nonlinear frequency drift of the probe laser \cite{Falke2011}. 

The largest uncertainty of the frequency standard results from the second-order Doppler shift caused by the residual secular and micromotion of the ion. The secular motion is determined from the observed carrier to sideband ratio as discussed in Ref.~\cite{Dube2013}. The temperature of the ion immediately after the cooling period is found to be 1.1~mK, which is close to the Doppler limit. However, the heating rate of $d\langle n\rangle/dt=190(60)$ quanta per second for the radial secular modes leads to an increased mean temperature of 2.0~mK during the interrogation. For the related frequency shift, we assume 50\% uncertainty since the ion temperature is not constantly monitored. Besides the shift caused by the thermal motion of the ion, excess micromotion caused by uncompensated stray fields can lead to additional shifts. As described in Ref.~\cite{Schneider2005}, we compensate the stray field by observing position changes of the ion while lowering the trap depth. From repeated compensation procedures with different initial conditions, we find that the stray field is compensated to better than $2.4$~V/m for each trap axis and we take the related maximum micromotion-induced Doppler shift as the uncertainty. The overall fractional Doppler shift is found to be $-3.7(2.1)\times 10^{-18}$. In addition to the Doppler shift, the residual interaction of the ion with the trapping field causes a Stark shift that can be calculated using $\Delta \alpha_s^\text{dc}$. In a trap with lower motional heating rates \cite{Nisbet-Jones2015}, ground state cooling can be advantageous, and with improved techniques for the cancellation of micromotion \cite{Keller2015}, we expect further reductions of these shifts and their uncertainties. Additionally, under these conditions a smaller fractional frequency instability can be achieved with longer interrogation times. 

We thank M.~Dole\u{z}al and P.~Balling for their work on the thermal analysis of the ion trap, H.~Lecher and F.~Brandt for the calibration of the power meter, and S.~Weyers for a critical reading of the manuscript. This work was supported by the European Metrology Research Programme (EMRP) in project SIB04. The EMRP is jointly funded by the EMRP participating countries within EURAMET and the European Union.

\end{document}